\documentclass[10pt]{iopart}
\usepackage{iopams}
\usepackage{setstack}
\usepackage{bm}
\usepackage{graphicx}
\usepackage{color}

\begin{document}

\title[Strong fields and neutral particle magnetic moment dynamics]{Strong fields and neutral particle magnetic moment dynamics}
\author{Martin Formanek, Stefan  Evans, Johann Rafelski,\\ Andrew Steinmetz, and Cheng-Tao Yang, } 
\address{Department of Physics, The University of Arizona, Tucson, AZ 85721, USA}

\begin{abstract}
Interaction of magnetic moment of point particles with external electromagnetic fields experiences unresolved theoretical and experimental discrepancies. In this work we point out several issues within the relativistic quantum mechanics and the QED and we describe effects related to a new covariant classical model of magnetic moment dynamics. Using this framework we explore the invariant acceleration experienced by neutral particles coupled to an external plane wave field through the magnetic moment: we study the case of ultra relativistic Dirac neutrinos with magnetic moment in the range of $10^{-11}$ to $10^{-20}$ $\mu_\mathrm{B}$; and we address the case of slowly moving neutrons. We explore how critical accelerations for neutrinos can be experimentally achieved in laser-pulse interactions. The radiation of accelerated neutrinos can serve as an important test distinguishing between Majorana and Dirac nature of neutrinos. 
\end{abstract}

\pacs{13.40.Em,06.30.Ka,41.75.Jv}

\vspace{2pc}
\noindent{\it Keywords}: magnetic moment, laser-driven acceleration, neutrino, neutron

\submitto{\PPCF} Accepted April 26, 2018

\ioptwocol

\section{Introduction}
The general consensus in theoretical physics is that the final word on classical Electrodynamics has not yet been said. More than a hundred and fifty years have passed since its original inception by Faraday, Maxwell and many others in the 19th century, and we still face unsolved conceptual problems of a fundamental nature. One of the most prominent issues of classical Electrodynamics is the problem of radiation reaction~\cite{Hadad:2010mt,Gralla:2009md,Spohn:2004ik}. 

Of comparable  relevance  is the incomplete understanding of the magnetic (Stern-Gerlach type) force, i.e. the interaction of the magnetic moment of a point particle with an external electro-magnetic (EM) field in both classical and quantum mechanics~\cite{Frenkel:1926zz, Foldy:1949wa}.  A related experimental discrepancy exists: as of July 2017 there is a  3.5 standard deviations difference between the calculated magnetic moment of the muon based on Standard Model QFT corrections and experimental measurements~\cite{Giusti:2017jof}. 

We report  on the recent progress in understanding the magnetic moment dynamics~\cite{Rafelski:2017hce}. Here we are interested in the dynamics of a neutral particle with non-zero magnetic moment placed in an external EM field. Any new magnetic moment physics is in this situation a first order effect. As an application of these considerations we describe how Dirac neutrinos could be studied experimentally, by exploiting their interaction with intense laser fields. We note another effort to improve  understanding of particle interaction with strong laser fields~\cite{Wen:2017zer}. Our work can also contribute to the study of plasma behavior influenced by external non-homogeneous fields. 
 
Before addressing the primary contents of this report we will first consider briefly the quantum physics of the magnetic moment in section~\ref{QM}, clarifying how the classical and quantum physics relate.  We summarize the insights of Ref.~\cite{Rafelski:2017hce} in  section~\ref{CovClas}, and we obtain the invariant acceleration acting on any particle in the plane wave field in section~\ref{a2}, before describing the physics of ultrarelativistic neutrinos in interaction with the plane wave field in section~\ref{sec:neutrino}.

\section{Relativistic Quantum Mechanics}\label{QM}
\subsection{Dynamical equations}
Every quantum particle should be described using three free parameters: its mass, its electric charge and its magnetic moment. However, the Dirac equation reduces the number of parameters to two, by predicting the magnetic moment $\mu = ge\hbar/2m$ with the gyromagnetic ration $g=2$. In reality, the effective $g$-factor is never exactly equal to two and in our effort to understand the dynamics of realistic particles we need to generalize our expressions to account for an anomalous magnetic moment with $a = g/2-1$. The deviancy can be small, such as in electrons and muons due to quantum electrodynamics effects, or large, such as in protons and neutrons due to their internal structures.  

The primary method of treating the anomalous magnetic dipole moment is by modification of the Dirac equation to include what is known as the Pauli term, containing the anomaly deviation $a \neq 0$ in the format
\begin{equation}\label{eq:DP}
(\gamma_\mu (i\hbar\partial^\mu - eA^\mu) - mc)\psi = a \frac{e\hbar}{4mc}\sigma_{\mu \nu} F^{\mu\nu} \psi\,.
\end{equation}
The main problem  with this approach is that the modified Dirac equation cannot be used to compute virtual processes, since the additional so called Pauli term diverges and requires counter terms.  

An alternative theoretical description of magnetic moment  first \lq\lq squares\rq\rq\ the Dirac equation, resulting in a second order formulation similar to the Klein-Gordon (KG) equation for spin 0 particles supplemented with the Pauli term
\begin{equation}\label{eq:KGP}
\left((i\hbar \partial_\mu - eA_\mu)^2 - \frac{ge\hbar}{4} \sigma^{\mu\nu}F_{\mu\nu} - m^2 c^2 \right) \psi = 0\,,
\end{equation}
where $\sigma^{\mu\nu} = \frac{i}{2}[\gamma^\mu,\gamma^\nu]$. A solution of the Dirac equation is also a solution of this KG-Pauli Eq.\,(\ref{eq:KGP}) once $g=2$ is chosen. The problem with KG-Pauli is that one must carefully analyze and understand the set of solutions of the higher order equation. 

The advantage of  KG-Pauli Eq.\,(\ref{eq:KGP}), compared to Dirac-Pauli Eq.\,(\ref{eq:DP}) is that we can choose an arbitrary value of the gyromagnetic factor $g$; if value $g=2$ \lq works\rq\, so will an arbitrary value. We emphasize that these  two quantum equations, the  Dirac-Pauli Eq.\,(\ref{eq:DP}) and the KG-Pauli Eq.\,(\ref{eq:KGP}),  are not equivalent and result in different physical behavior. Thus experiment will determine which form corresponds to the quantum physics of e.g. bound states in hydrogen-like Coulomb potential. We will return to this matter under seperate cover.
   
\subsection{Magnetic moment in QED}
In principle quantum electrodynamics is formulated around a Dirac particle with $g=2$ with modifications arising in the context of a perturbative expansion leading to the evaluation of the actual magnetic moment, i.e. $g\ne 2$ of the electron in perturbative series that today requires in precision study also the consideration of strong interactions and vacuum structure.  This approach masks the opportunity to use the actual particle magnetic moment for particles responsible for the vacuum properties such as in vacuum polarization.
 
The study of the vacuum response to external fields has a long and distinguished history that spans over 80 years, starting with computation of the lowest order effect by Uehling~\cite{Uehling:1935uj} in 1935 and the development of the nonperturbative Euler-Heisenberg (EH) effective action characterizing all the physical phenomena present in  constant fields, including the decay of the field into electron positron pairs~\cite{Heisenberg:1935qt}. These studies introduce counter terms which served as predecessors to the full quantum field theoretical charge renormalization scheme. This effective action was revisited in a field theoretical context by Schwinger, which extended these considerations to include  a demonstration of transparency of the vacuum to a single electromagnetic plane wave~\cite{Schwinger:1951nm}. However, all these consideration required particles to have the Dirac value of magnetic moment $g=2$. 

When $g\neq 2$ is introduced, a modification of the analytical form of the effective EH action is discovered~\cite{Labun:2012jf,Rafelski:2012ui} and further non-trivial modifications in the vacuum structure arise~\cite{AngelesMartinez:2011nt,VaqueraAraujo:2012qa}. A solution to the previously divergent result for effective action with $|g|>2$ was obtained~\cite{Rafelski:2012ui}. Similarly  the modification of the vacuum polarization was found~\cite{AngelesMartinez:2011nt}
\begin{eqnarray}\label{eq:VP}
\fl \pi(q^2)=-\frac {e^2}{12\pi^2}&
\left(\frac{3}{8} g^2  -\frac{1}{2}\left(1-\frac{4m^2}{q^2}\right)\right)
\times \nonumber \\
&\Big[\frac13+\int_0^1dx\ln\Big(1-\frac{q^2}{m^2}x(1-x)\Big)\Big]\;.
\end{eqnarray} 
The coefficient in Eq.\,(\ref{eq:VP}) shows explicitly all three parameters of a particle: its magnetic moment in form of $g$, its charge $e$ and its mass $m$. One can easily recombine terms to show dependence on the magnetic anomaly $a=g/2-1$. This form demonstrates that in perturbative QED expansion, the magnetic moment dependence arises from the higher order QED  vacuum polarization tensor (the photon line crossing the loop) contributing. This format hides the appearance of the actual particle magnetic moment in the vacuum polarization as is seen in Eq.\,(\ref{eq:VP}). We will return to the question how magnetic moment is renormalized under separate cover. 

Once we recognize the dependence of vacuum polarization on magnetic moment and the dependence of EH effective action on magnetic moment one must further revisit Schwinger\lq s proof of vacuum transparency to a single plane wave  for $g\neq2$. 

\section{Magnetic moment in classical theory}\label{CovClas}
There are two models which describe the magnetic moment of a point particle. The \lq Amperian\rq\  Model approximates the particle magnetic moment by a current loop which leads to a force
\begin{equation}
\bm{F}_{ASG} = \nabla(\bm{\mu}\cdot \mathcal{\bm{B}})\;,
\end{equation}
where $\bm{\mu}$ is the magnetic moment of the particle and $\mathcal{B}$ is magnetic field. On the other hand the \lq Gilbertian\rq\  Model creates a magnetic dipole, consisting of two hypothetical monopoles, and leads to a different expression
\begin{equation}
\bm{F}_{GSG} = (\bm{\mu} \cdot \nabla) \mathcal{\bm{B}}\;.
\end{equation}
We expect that there should be a way  to reconcile these classical models and  to create a covariant description of the dynamics for both particle 4-velocity $u^\mu$ and spin $s^\mu$, which would unite these two approaches. There have been efforts to do so - the first covariant model was created by Frenkel~\cite{Frenkel:1926zz,Frenkel:1926spin}. This model is based on classical arguments starting with the principle of least action and couples back the spin motion of the particle with the particle motion. 

Another method of approach begins with relativistic quantum Dirac theory which naturally incorporates description of the spin behavior (although  $g=2$  strictly)  and finding an appropriate classical limit should yield a full classical description of the particle behavior. The most important example of such an approach is the Foldy-Wouthuysen transformation~\cite{Foldy:1949wa} 

Both of these approaches  predict different behavior in the external EM field and can be distinguished experimentally as was explored in the article~\cite{Wen:2017zer}. We learn from this work that ultra-intense laser pulses are especially suitable for investigating the viability of such models. 
  
As presented in the work~\cite{Rafelski:2017hce}, the spin of a particle should not be its quantum property but rather a classical characteristic similar to the particle\lq s mass. Both of these are eigenvalues of Casimir operators of the Poincar\' e group of space-time symmetry transformations, whose values describe a representation of this group for a given particle. This insight allowed us to create a new covariant description of the spin dynamics of particles~\cite{Rafelski:2017hce} which has the form
\begin{equation}\label{eq:udynam}
\dot{u}^\mu = \frac{1}{m} (q F^{\mu \nu} - s \cdot \partial F^{*\mu\nu}d)u_\nu\;,
\end{equation}
\begin{eqnarray}
\fl \dot{s}^\mu = \frac{1+\widetilde{a}}{m}&\left(qF^{\mu\nu} -\frac{1+\widetilde{b}}{1+\widetilde{a}}s\cdot\partial F^{*\mu\nu}d\right)s_\nu \nonumber\\
&- \widetilde{a}\frac{u^\mu}{mc^2}\left(u\cdot \left( qF - \frac{\widetilde{b}}{\widetilde{a}}  s \cdot\partial F^{*}d\right) \cdot s\right)\label{eq:sdynam}\;,
\end{eqnarray}
where  $\widetilde{a}$ and $\widetilde{b}$ are arbitrary constants. We explicitly distinguish between particle charge $q$ and elementary magnetic dipole charge $d$, which is used to convert the spin of a particle $\bm{s}$ to the magnetic moment $\bm{\mu}$ as $c|\bm{s}|d \equiv |\bm{\mu}|$. Finally, the dual EM tensor reads $F^{*\mu\nu} = \epsilon_{\mu\nu\alpha\beta}F^{\alpha\beta}/2$,
with the fully antisymmetric tensor defined as $\epsilon_{0123} \equiv +1$ (beware of a sign if contravariant indices are used). 

We see that the equation of motion Eq.\,(\ref{eq:udynam}) of the particle  depends explicitly on the spin dynamics Eq.\,(\ref{eq:sdynam}) through the spin 4-vector $s^\mu(\tau)$, thus generating covariant generalization of the Lorentz force to include Stern-Gerlach force. For particles with zero magnetic moment $d = 0$ these dynamical equations reduce to Thomas-Bargmann-Michel-Telegdi (TBMT) equations~\cite{Thomas:1926dy,Bargmann:1959gz} with $\widetilde{a} = a$. TBMT equations are widely used to model particle dynamics in external fields and yet these do not contain coupling of the spin to the particle motion. 

On the other hand, we can also explore the other limit:  the dynamics of neutral particles $q = 0$ with magnetic moment $d \neq 0$ in external fields. Equations\,(\ref{eq:udynam}),\,(\ref{eq:sdynam}) become only functions of parameter $\widetilde{b}$, which we will further explore in section~\ref{sec:neutrino}.

To conclude this short overview of the results obtained in Ref.\,\cite{Rafelski:2017hce} we note that the two forms of the force, the Amperian and the Gilbertian, were shown to be equivalent. Thus a consistent theoretical framework now exists for exploring the dynamics of a magnetic moment in external fields. 

\section{ Dynamics of particles in a plane wave field}\label{a2}
\subsection{Invariant acceleration}
The generalized Lorentz force equation reads~\cite{Rafelski:2017hce}, see Eq.\,(\ref{eq:udynam})
\begin{equation}\label{eq:genlorentz}
\dot{u}^\mu = \frac{1}{m}\widetilde{F}^{\mu\nu}u_\nu, \quad \widetilde{F}^{\mu\nu} \equiv q F^{\mu\nu} - s \cdot \partial F^{*\mu\nu}d\;. 
\end{equation}
Imagine a point particle with both electric charge and magnetic moment in the plane wave field given by expression
\begin{equation}\label{eq:properties}
A^\mu(\xi) = \mathcal{A}_0 \varepsilon^\mu f(\xi), \quad \xi = k \cdot x, \quad k \cdot \varepsilon = 0, \quad k^2=0\;,
\end{equation}
where $k^\mu$ is a wave vector of the plane wave; $\varepsilon^\mu$ its polarization; $\xi$ phase; and $\mathcal{A}_0$ amplitude. $f(\xi)$ is a function characterizing the laser pulse. Just the formula for the dynamics of the 4-velocity Eq.\,(\ref{eq:genlorentz}) alone is sufficient to obtain an expression for invariant acceleration in the plane wave field. In this case the generalized EM tensor reads
\begin{eqnarray}
\fl \widetilde{F}^{\mu\nu} = &\mathcal{A}_0q(k^\mu\varepsilon^\nu - k^\nu \varepsilon^\mu)f^\prime (\xi) \nonumber \\
&- \mathcal{A}_0f^{\prime \prime}(\xi)(k\cdot s)\epsilon^{\mu\nu\alpha\beta}k_\alpha \varepsilon_\beta d\;,
\end{eqnarray}
where primes denote derivatives of the pulse function $f(\xi)$ with respect to its phase. If we multiply this expression with $k^\mu$ we get zero because of the identities in Eq.\,(\ref{eq:properties}). Then  Eq.\,(\ref{eq:genlorentz}) implies that 
\begin{equation}\label{eq:integral}
k \cdot \dot{u} = 0, \quad \Rightarrow \quad k \cdot u = k \cdot u(0)\;,
\end{equation}
is an integral of motion. We can obtain the invariant acceleration by squaring the expression  Eq.\,(\ref{eq:genlorentz}), which can be evaluated using  Eq.\,(\ref{eq:properties}), antisymmetric properties of $\epsilon$, our integral of motion  Eq.\,(\ref{eq:integral}), and contraction identity
\begin{equation}
\epsilon^{\mu\nu\alpha\beta}\epsilon_{\mu\rho\gamma\delta} = - \delta^{\nu\alpha\beta}_{\rho\gamma\delta}\;,
\end{equation}
which is a generalized Kronecker delta. The final result is
\begin{equation}\label{eq:invaracc}
\dot{u}^2 = - \frac{\mathcal{A}_0^2}{m^2}\left[q^2f^\prime (\xi)^2 + (k\cdot s)^2 f^{\prime \prime} (\xi)^2 d^2 \right] (k\cdot u(0))^2\;.
\end{equation} 
The cross term vanishes because the force due to particle electric charge and magnetic moment are orthogonal for a plane wave field. The only unknown in this expression is the product $(k \cdot s(\tau))$, which is still a function of proper time.  

\subsection{ Neutral particle dynamics}
As explained in the reference \cite{Rafelski:2017hce} the  torque Eq.\, (\ref{eq:sdynam}) is constructed to be compatible with the force Eq.\,(\ref{eq:genlorentz}). For neutral particles we require in addition as did Ref.\cite{Morley:2014pja} that torque involves full magnetic moment; that is, for the particle at rest in the laboratory frame we have the  torque $\propto \bm{\mu} \times \bm{B}$. Restating the force equation for neutral particles the two   dynamical equations thus are
\begin{equation}\label{eq:udynam}
\dot{u}^\mu = - s \cdot \partial F^{*\mu\nu} u_\nu \frac{d}{m}\;,
\end{equation}
\begin{equation}
\dot{s}^\mu = cd \left(F^{\mu\nu}s_\nu - \frac{u^\mu}{c^2}(u\cdot F \cdot s)\right)- s\cdot\partial F^{*\mu\nu}s_\nu \frac{d}{m}\;.
\end{equation}
The full analytical solution of these equations are in preparation for publication under separate cover. Here of importance is the solution for the projection of spin on the wave vector of the laser 
\begin{eqnarray}
\fl k \cdot s(\tau) &= k \cdot s(0) \cos[\mathcal{A}_0 d (f(\xi(\tau)) - f(\xi_0))] \nonumber\\
&- \frac{W}{c} \sin[\mathcal{A}_0 d (f(\xi(\tau)) - f(\xi_0))] \label{eq:spinprec}\;,
\end{eqnarray} 
where $W$ is determined by initial conditions
\begin{equation}\label{eq:w}
W \equiv [(k \cdot u(0))(\varepsilon \cdot s(0)) - (\varepsilon \cdot u(0))(k \cdot s(0))]\;.
\end{equation}
It is very important to know $k \cdot s(\tau)$  because the invariant acceleration of the particle, obtained by squaring Eq.\,(\ref{eq:udynam}), is
\begin{equation}\label{eq:neutinvacc}
\dot{u}^2(\tau) = - (k \cdot s(\tau))^2 (k \cdot u(0))^2 f^{\prime\prime}(\xi)^2 \frac{\mathcal{A}_0^2d^2}{m^2}\;.
\end{equation}
The invariant acceleration therefore depends on the products $(k \cdot u(0))$ and $(k \cdot s(\tau))$. The first one is a Doppler shifted laser frequency as seen by the particle being hit by the laser pulse. In the laboratory frame with 
\begin{equation}
u^\mu(0) = \gamma_0 c (1, \bm{\beta}_0)\;,\ \; k^\mu = \omega(1,\hat{\bm{k}})/c\;, \ \; \epsilon^\mu = (0, \hat{\pmb{\varepsilon}})\;,
\end{equation}
we can write
\begin{equation}\label{eq:doppler}
k \cdot u(0) = \gamma_0 (1 - \hat{\bm{k}} \cdot \bm{\beta}_0)\omega\;.
\end{equation}
To evaluate Eq.\,(\ref{eq:neutinvacc}) we further need  $(k \cdot s(0))$, denoting the initial alignment of the particle spin and the wave vector. Since $u \cdot s = 0$, the initial spin 4-vector in the laboratory frame reads
\begin{equation}
s^\mu_L(0) = (\bm{\beta}_0 \cdot \bm{s}_{0L}, \bm{s}_{0L})\;, 
\end{equation}
\begin{equation}
\bm{s}_{0L} = \bm{s}_0 + \frac{\gamma_0 - 1}{\beta_0^2}(\bm{\beta}_0 \cdot \bm{s}_0) \bm{\beta}_0\;,  
\end{equation}
where   $\bm{s}_{0L}$ is the Lorentz transform of the initial spin of the particle  $\bm{s}_0$ given in its rest frame. Therefore
\begin{equation}
\frac{c}{\omega}k \cdot s(0) =  \gamma_0 (\bm{\beta}_0 \cdot \bm{s}_0) - (\gamma_0 - 1)(\hat{\bm{\beta}}_0 \cdot \bm{s}_0)(\hat{\bm{\beta}}_0 \cdot \hat{\bm{k}}) - \hat{\bm{k}} \cdot \bm{s}_0 \;.
\end{equation}
For the particle beam pointing against the laser pulse $\hat{\bm{\beta}}_0 \cdot \hat{\bm{k}} = -1$, we pick up a factor of $\gamma_0$
\begin{equation}\label{eq:ks0}
\frac{c}{\omega}k \cdot s(0) = \gamma_0(\beta_0 + 1)(\hat{\bm{\beta}}_0 \cdot \bm{s}_0) - (\hat{\bm{\beta}}_0 \cdot \bm{s}_0) -  \hat{\bm{k}} \cdot \bm{s}_0  \;.
\end{equation}
$k \cdot s(0)$ factor plays an important role in the ultrarelativistic interactions discussed in the following section.

Finally, the combination of the initial conditions Eq.\,(\ref{eq:w}) evaluated in the laboratory frame reads
\begin{eqnarray}\label{eq:wc}
\fl \frac{W}{c} = \gamma_0 \frac{\omega}{c}\big[&-\hat{\pmb{\varepsilon}} \cdot \bm{s}_0 + \left( 1 - \frac{1}{\gamma_0}\right)(\hat{\bm{\beta}}_0 \cdot \bm{s}_0)(\hat{\bm{\beta}}_0 \cdot \hat{\pmb{\varepsilon}}) \nonumber\\
 &+ (\hat{\bm{k}} \cdot \bm{\beta}_0)(\hat{\pmb{\varepsilon}} \cdot \bm{s}_0) - (\bm{\beta}_0 \cdot \hat{\pmb{\varepsilon}})(\hat{\bm{k}} \cdot \bm{s}_0) \big]\;,
\end{eqnarray}
which is also proportional to only one (in general highly relativistic) $\gamma_0$ factor.

\section{Neutrino acceleration {\small (ultrarelativistic limit)}}\label{sec:neutrino}
As discussed in preceding sections we are especially interested in the case of charge neutral particles in the external EM fields. The most prominent examples of such particles are neutrons and neutrinos. In the absence of the classical Lorentz force the particle dynamics is governed by  spin effects and can directly be used to measure the related properties of particles. 

The interaction of neutrinos with a laser field was studied previously \cite{Meuren:2015iha} as a higher order scattering effect, but in the framework we developed~\cite{Rafelski:2017hce} neutrinos couple with external fields via magnetic moment directly. 

We recall that by symmetry arguments only the Dirac neutrino can have a magnetic moment: in essence this is because the Majorana neutrino is the antiparticle of itself and thus under EM interactions must be neutral in both charge and magnetic moment. A very significant effort is underway to discover the double beta-decay~\cite{Alduino:2017ehq} that could demonstrate that the neutrino is of the Majorana type. However, one can question if a nil result would mean that the neutrino is a Dirac neutrino~\cite{Hirsch:2017col}. We believe that the measurement of neutrino interactions with an external field via its magnetic moment would demonstrate that the neutrino is of the Dirac type. Our objective in the following is to show that we not only can expect observable effects when relativistic neutrinos interact with an intense EM plane wave pulse, but that a measurement of the magnetic moment of the neutrino should be possible.

\subsection{Magnetic moment of the neutrino}
The dipole magnetic moment is a well studied electromagnetic property of the Dirac neutrino. A minimal extension of the Standard Model with non-zero Dirac neutrino masses places a lower bound on the magnetic moment of the neutrino mass eigenstate $\nu_i$ proportional to its mass $m_i$ and reads~\cite{Fujikawa:1980yx}
\begin{equation}\label{eq:low}
\mu_{i}=\frac{3\,G_F\,m_e\,m_i}{4\,\sqrt{2}\,\pi^2}\mu_B=3.2\times10^{-19}\left(\frac{m_i}{\mathrm{eV}}\right)\mu_B,
\end{equation}
where $\mu_B = e\hbar/2m_e$ is the Bohr magneton. This value is several orders of magnitude smaller than the present experimental upper bound~\cite{Patrignani:2016xqp} 
\begin{equation}\label{eq:up}
\mu_\nu < 2.9\times 10^{-11} \mu_B\;.
\end{equation}

\subsection{Neutrino acceleration in the external field}
We consider a beam of neutrinos with $E_\nu \simeq 20$ GeV. This energy of neutrinos is currently accessible, for example the OPERA experiment used 17 GeV neutrinos produced at CERN~\cite{Duchesneau:2017tpi}. For the rest mass of neutrinos we take $m_\nu = 0.2$ eV and laser source with photon energy $E_\gamma = 1$ eV. The de Broglie wavelength for such neutrinos compared to the wavelength of the laser light is
\begin{equation}
\frac{\lambda_\nu}{\lambda_\gamma} = \frac{E_\gamma}{E_\nu(kin)} \approx \frac{E_\gamma}{E_\nu} \approx 5 \times 10^{-11}\;, 
\end{equation}
where we neglected the mass of the neutrinos compared to their energy. This justifies the classical treatment because the wavelength of the 1 eV laser light is 11 orders of magnitude larger than the wavelength of the 20 GeV neutrinos, therefore the quantum wave character of neutrinos will be invisible.  
The amplitude $\mathcal{A}_0$ of the laser field vector potential can be expressed in terms of the dimensionless normalized amplitude $a_0$ as 
\begin{equation}\label{eq:hata}
\mathcal{A}_0 = \frac{m_ec}{e} a_0\;.
\end{equation}
The current state of the art for laser systems is $a_0 \sim 10^2$. The elementary dipole charge of the neutrino can be rewritten using the neutrino magnetic moment in units of Bohr magneton as
\begin{equation}\label{eq:first}
d = \frac{e}{m_e c} \mu[\mu_B]\;.
\end{equation}
This makes the relevant product
\begin{equation}\label{eq:A0d}
\mathcal{A}_0 d = a_0 \mu_\nu[\mu_B] \approx 10^{-9} - 10^{-11}
\end{equation}
for state of the art laser systems and possible range of values for neutrino magnetic moment Eqs.\,(\ref{eq:low},\,\ref{eq:up}). From the Eq.\,(\ref{eq:wc}) we get ultrarelativistic limit for $W/c$ term and using (Eq.\,(\ref{eq:ks0})) ultrarelativistic limit for product $k \cdot s(0)$
\begin{equation}
\frac{W}{c} \sim \frac{\gamma_0 \hbar \omega}{c}\;, \qquad k \cdot s(0) \sim \frac{\gamma_0 \hbar \omega}{c}\;.
\end{equation}
This means that for our extremely small  $\mathcal{A}_0 d$ (Eq.\,(\ref{eq:A0d})) we can see from Eq.\,(\ref{eq:spinprec}) that there is no  (neutrino) spin precession and 
\begin{equation}
k \cdot s(\tau) \approx k \cdot s(0)\;,
\end{equation}
with a very high precision. 

Equation\,(\ref{eq:neutinvacc}) allows us to evaluate the invariant acceleration which the 20 GeV neutrino experiences in the external plane wave field
\begin{equation}\label{eq:neutrinoacc}
\sqrt{\dot{u}^2} \approx \left| (k \cdot s(0))(k\cdot u(0)) f^{\prime \prime} (\xi)\frac{\mathcal{A}_0d}{m_\nu}\right|\;.
\end{equation}
We turn now to  estimate individual terms:\\
i)  The Doppler shifted frequency $(k \cdot u(0))$ is given in the laboratory frame by the formula (\ref{eq:doppler}) and for the ultra relativistic neutrinos with velocity $\bm{\beta}_0$ oriented against the laser beam propagation direction $\hat{\bm{k}}$ we can write
\begin{equation}
k \cdot u(0) = \gamma_0 (1 - \hat{\bm{k}} \cdot \bm{\beta}_0) \omega \approx  2\gamma_0 \omega \approx 2\frac{E_\nu E_\gamma}{m_\nu c^2 \hbar}\;.
\end{equation}
ii) The product $k \cdot s(0)$ (Eq.\,(\ref{eq:ks0})) is in the ultrarelativistic case proportional to
\begin{equation}
k \cdot s(0) \approx \frac{\gamma_0 \hbar\omega}{c} \approx \frac{E_\nu}{m_\nu c^2}\frac{E_\gamma}{c}\;.
\end{equation}
iii) Finally, we want to write the result in the units of critical acceleration for the neutrino which is
\begin{equation}\label{eq:last}
a_c = \frac{m_\nu c^3}{\hbar}\;.
\end{equation}
Substituting all terms in equations,(\ref{eq:hata}-\ref{eq:last}) into (\ref{eq:neutrinoacc}) yields an expression for the acceleration
\begin{eqnarray}
\sqrt{\dot{u}^2}[a_c] \approx a_0 f^{\prime \prime} (\xi) \frac{(E_\nu[eV])^2 (E_\gamma[eV])^2}{(m_\nu[eV])^4} \mu_\nu[\mu_B]\;. 
\end{eqnarray}
For our 20 GeV neutrinos we see that the critical acceleration can be achieved in the whole range of magnetic moment that Dirac neutrinos could have $\mu_\nu \in (10^{-11} - 10^{-20})\mu_B$ for corresponding laser pulse parameters in the range 
\begin{equation}
a_0 f^{\prime \prime} (\xi) \in (10^{-13} - 10^{-4})\;.
\end{equation}
The state of the art laser systems have dimensionless normalized amplitude $a_0$, Eq.\,(\ref{eq:hata}) on the order of $10^2$. Even the second derivative of the laser pulse function can be high, because typically we get $f(\xi)$ as a product of oscillating function $\sin(\xi)$ and envelope $g(\xi)$ which has a second derivative
\begin{eqnarray}
\fl f''(\xi) &= (\sin(\xi)g(\xi))'' \nonumber \\
&= -\sin(\xi)g(\xi) + 2 \cos(\xi)g'(\xi) + \sin(\xi)g''(\xi) \;.
\end{eqnarray}
The dominant term that we can exploit is the first derivative of the envelope function which can be very high on the front of the pulses with high contrast ratio. For example if the intensity of the light drops by 99\% from the maximum on the distance of half wavelength (therefore field amplitude drops by 90\% on the same distance) we get $g'(\xi) = 0.9/\pi ~ \sim 10^{-1}$. Thus we believe that critical neutrino acceleration can be achieved for the whole range of permissible neutrino magnetic moment with accessible laser systems. 

Relativistic high intensity neutrino beams are available, and continue to be developed, at particle accelerators (CERN, Fermilab) for neutrino oscillation experiments and related \lq intensity frontier\rq\ research. The typical energy of a high intensity $\nu,\,\bar\nu$-beam is at 10-20 GeV level, but a beam-dump sourced beam at CERN-LHC would produce neutrinos with 100 times higher energy. This high-energy beam of neutrinos responds by a factor $\gamma_0^2$ in our favor. In comparison to accelerator sourced neutrinos, the highest natural $\nu$-flux on Earth is at 0.6--1 MeV from $pp$-solar fusion chains. Interactions with the laser light at this energy would be suppressed by a factor $10^8$ compared to the 10 GeV neutrino beam, but the solar source \lq shines\rq\ with 100\%  duty cycle tracking sky location of the Sun also across the Earth. This shows that before an experiment can be realized, prioritization and optimization between the intensity of the laser light, the accessible energy of neutrinos, and the luminosity of the neutrino flux have to be studied in order to select an optimal experimental environment

\subsection{Neutrino radiation}
There are multiple ways how an accelerated neutrino can radiate. It certainly produces magnetic dipole electromagnetic radiation as discussed in Refs.\cite{Morley:2014pja,jackson}.

At 20 GeV energy it is even possible that the neutrino will emit (virtual) electro-weak bosons $W^\pm$ and $Z^0$, which will decay into  relatively high 10-GeV energy scale, and thus more easily observable, either dilepton pairs, and/or hadronic showers (hadronic decay  of $Z^0,W^\pm$). Thus with some probability shooting a laser pulse onto an incoming 20 GeV neutrino beam may catalyze  GeV scale particle production, a process that would be hard to interpret otherwise.

While experiments seeking double-$\beta$-decay of Majorana neutrinos are underway, an experiment seeking evidence for Dirac neutrino has not been available before. The possible ultra-intense  laser pulse catalysis of radiation by an ultra-relativistic neutrino provides this opportunity for the first time. Therefore these processes will be subject to future study. Aside demonstrating possible Dirac nature of the neutrino, such experiments would  provide vital information about the neutrino magnetic moment and mass.

\section{Neutrons}
\subsection{Neutron acceleration}
Given that a neutron is about  $5 \times 10^{9}$ heavier compared to a neutrino one cannot expect a Lorentz-factor $\gamma_0=E_n/m_nc^2$ that is anywhere near to the value $10^{11}$ that makes neutrino magnetic interactions with the external field strong. Even so, we note that  iThemba LABS can produce neutrons with kinetic energy of $~200$ MeV~\cite{Mosconi:2010char}, which corresponds to $E_n \approx 960$ MeV. This still places their dynamics into a classical regime, since $\lambda_n/\lambda_\gamma ~\approx 5\times 10^{-9}$ for 1 eV laser photons.

Even though the magnitude of the magnetic moment for the neutron is several orders of magnitude larger $|\mu_n| = 1\times 10^{-3} \mu_B$ the neutrons are $10^{9}$ times heavier and in conclusion we would require the product $a_0 f^{\prime \prime}(\xi)$ to be as high as $10^{23}$ in order to achieve critical accelerations which is definitely not currently accessible. On the other hand the neutron-external magnetic field interaction is appreciable and has been used to keep a neutron beam in a storage ring~\cite{Paul:1989hn}. The EM plane wave  interaction with neutrons introduces a novel method of neutron motion and spin motion controle.

\subsection{Polarization of non-relativistic neutrons}
We need the neutrons to move slowly enough so that we can consider the non-relativistic limit, but not so slowly that we have to take into account quantum mechanical effects. For example, slow neutrons at 10 eV have $\beta_0 \sim 10^{-4}$ and $\lambda_n / \lambda_\gamma \sim 10^{-6}$ for a 1 eV laser source, which still puts them well into the classical region.

This time the product $\mathcal{A}_0 d$ which governs the spin precession is for the state of the art laser appreciable $\mathcal{A}_0 d \approx 10^{-1}$ which makes both terms in the spin precession Eq.\,(\ref{eq:spinprec}) relevant and spin of neutron indeed rotates in the external laser field.

The estimate of the $k \cdot s(0)$ term (Eq.\,(\ref{eq:ks0})) in the zeroth order of $\beta_0$ is
\begin{equation}
k \cdot s(0) \approx - \frac{\omega}{c}\hat{\bm{k}} \cdot \bm{s}_0\;, 
\end{equation}
and the term $W/c$ (Eq.\,(\ref{eq:wc})) in the zeroth order of $\beta_0$ reads
\begin{equation}
\frac{W}{c} \approx - \frac{\omega}{c} \hat{\pmb{\varepsilon}} \cdot \bm{s}_0\;.
\end{equation} 
This means that the spin precession equation, Eq. (\ref{eq:spinprec}), reduces to
\begin{eqnarray}\label{eq:spinprecNR}
\fl \hat{\bm{k}} \cdot \bm{s}(t) &\approx \hat{\bm{k}} \cdot \bm{s}_0 \cos[\mathcal{A}_0 d (f(\xi(t)) - f(\xi_0))] \nonumber\\
&- \hat{\pmb{\varepsilon}} \cdot \bm{s}_0 \sin[\mathcal{A}_0 d (f(\xi(t)) - f(\xi_0))]\;. 
\end{eqnarray} 

We see in Eq.\,(\ref{eq:spinprecNR}) that just as in the relativistic result Eq.\,(\ref{eq:spinprec}), in the non-relativistic limit the spin precesses with $\mathcal{A}_0 d  f(\xi(t))$. However, unlike for neutrinos, given the large magnetic moment of neutrons, the spin precession can be significant. The spin projection oscillates between initial alignments with direction of plane wave propagation and against  the polarization vector (and vice-verse). This is in agreement with the expectation based on non-relativistic torque action, which we will further discuss elsewhere.

\subsection{Neutron lifespan} 
Closing the discussion of neutron dynamics we draw attention  to the recent recognition that neutron decay anomaly, i.e. the lifespan inconsistency between \lq in bottle\rq, and \lq in flight\rq\ measurements, could be related to an unknown dark matter decay of the neutron~\cite{Fornal:2018eol}. 

We have explored the question if this inconsistency could be due to the neutron lifespan being affected by the strong field environment accompanying the \lq in flight\rq\ type measurement experiments \cite{Nico:2004ie}. We were considering the modification of the proper time by the strong field. Since it is hard to accelerate neutrons using their magnetic moment we did not identify an effect. However, this lifespan discrepancy and associated presence of strong fields remains a topic deserving further theoretical and, in the context of laser strong fields, novel experimental investigation employing  i.g. neutrons kept in a storage ring~\cite{Paul:1989hn} accompanied by a EM plane wave.

\section{Conclusions} 
The novel domain of EM magnetic moment interactions in external fields which has been recently formulated also holds  promise to enhance the understanding of physics of plasmas. In this paper we focused on the dynamics on neutral particles, namely neutrinos and neutrons. The purpose of this paper was to introduce new particle physics opportunities present in the ultra intense laser physics frontier. Among results we have obtained is for example that (ultra-relativistic) neutrinos embedded in ultra-strong high contrast laser pulses are not subjected to any appreciable spin precession unlike neutrons for which spin dynamics becomes important. 

The relevant   Eq.\,(\ref{eq:spinprec}) and, respectively,   Eq.\,(\ref{eq:spinprecNR}) depend alone on the behavior related to TBMT torque dynamics, section \ref{CovClas}. However, a prior study of covariant neutron (spin) dynamics in the presence of a EM plane wave  is not known to us and we believe that these results are presented here for the first time.

We have proposed exploration of laser pulse interaction with ultrarelativistic neutrinos. As our discussion shows the ultra large neutrino Lorentz-$\gamma_0$ factor enhances the interaction strength with the external field opening opportunity to revolutionize the study of  physical properties  of the Dirac neutrinos as both the mass and the magnetic moment  could be studied through the magnetic moment EM radiation and/or $W, Z$ radiation decay channels. 

Here it is important to realize that only a positive outcome of the double-$\beta$ decay experiment proves that neutrino is a Majorana particle; in the absence of a result a  complementary experiment aiming to recognize Dirac neutrino magnetic moment would serve as an important test which could resolve the question whether the neutrino is a Dirac or a Majorana particle.

\end{document}